\documentclass{myaa}
\usepackage{graphicx}
\def\kms{km\,s$^{-1}$}
\def\hb{H$\beta$}
\def\ha{H$\alpha$}
\def\nii{\lbrack\ion{N}{ii}\rbrack}
\def\oiii{\lbrack\ion{O}{iii}\rbrack}
\def\oii{\lbrack\ion{O}{ii}\rbrack}

\def\sii{\lbrack\ion{S}{ii}\rbrack}

\def\oi{\lbrack\ion{O}{i}\rbrack}
\def\roiii{$\lambda$5007$/$H$\beta$}

\def\rnii{$\lambda$6584$/$H$\alpha$}
\def\roi{$\lambda$6300$/$H$\alpha$}
\def\caii{\ion{Ca}{ii}}
\def\hii{\ion{H}{ii}}
\def\hi{\ion{H}{i}}

\def\ergs{erg\,s$^{-1}$}
\def\a843{$\alpha^{8}_{43}$}
\def\apj{ApJ }

\def\beq{\begin{equation}}
\def\eeq{\end{equation}}

\begin{document}
    \title{The Nuclear Region of \\
Low Luminosity Flat Radio Spectrum Sources\thanks{Based on observations
collected at the Multiple Mirror Telescope on Mt. Hopkins.}}

\subtitle{II. Emission-Line Spectra}

    \author{A.\,C. Gon\c{c}alves\inst{1,2} \and M. Serote Roos\inst{1}}

    \offprints{A.\,C. Gon\c{c}alves}

\institute{Centro de Astronomia e Astrof\'{\i}sica da Universidade de Lisboa,
Observat\'orio Astron\'omico de Lisboa,  Tapada da Ajuda, P-1349-018 Lisboa,
Portugal \\
\email{adarbon@oal.ul.pt, serote@oal.ul.pt}
\and
LUTH, Observatoire de Paris--Meudon,  5 Place Jules Janssen, F-92195 Meudon
Cedex, France\\
\email{anabela.goncalves@obspm.fr}}

  \date{Received: 8 January 2003 / Accepted 4 September 2003}

% ---------------------------------------------------------------------
\abstract{
We report on the spectroscopic study of 19 low luminosity Flat 
Radio Spectrum (LL FRS) sources selected from March\~a's et al. 
(1996) 200~mJy sample. In the optical, these objects are mainly 
dominated by the host galaxy starlight. 
After correcting the data for this effect, we obtain a new 
set of spectra clearly displaying weak emission lines;  
%This paper is dedicated to their analysis. 
such features carry valuable information concerning the 
excitation mecha\-nisms at work in the nuclear regions of 
LL FRS sources. We have used a special routine to model 
the spectra and assess the intensities and velocities of 
the emission lines; we have analyzed the results in terms of diagnostic 
diagrams. Our analysis shows that 79\% of the studied 
objects harbour a Low Ionization Nuclear Emission-line 
Region (or LINER) whose contribution was  swamped by the 
host galaxy starlight. The remaining objects display a 
higher ionization spectrum, more typical of Seyferts; due to 
the poor quality of the spectra, it was not possible to identify  
any possible large Balmer components.  
The fact that we observe a LINER-type spectrum 
in LL FRS sources supports the idea that some of these objects 
could be undergoing an ADAF phase; in addition, such a low ionization 
emission-line spectrum is in agreement with the black hole mass 
values and sub-Eddington accretion rates published for some FRS  
sources. 
\keywords{Galaxies: active  -- Galaxies: nuclei -- Galaxies:
          BL~Lacertae objects: general -- Galaxies: general}
}
  \titlerunning{The Nuclear Region of LL FRS Sources: Emission-Line Spectra}
  \authorrunning{A.\/C.\,Gon\c{c}alves  \& M.\,Serote Roos}
  \maketitle
%----------------------------------------------------------------------

\section{Introduction}
Low frequency surveys of extragalactic radio sources have 
observed predominantly extended double-lobed radio-galaxies,
which display a steep radio spectrum. Surveys carried out at
higher frequencies  found a larger number of flat radio spectrum
(FRS) sources, i.e. sources displaying hard radio spectra 
and usually characterized by an unresolved, compact radio-core. 

Amongst flat radio spectrum sources, we find  BL~Lacertae objects
(BL Lacs). BL Lacs are a class of Active Galactic Nuclei (AGN)
characterized by extreme properties, such as high X-ray and
$\gamma$-ray luminosities, relatively high optical and radio
polarization, and strong and rapid variability at radio, optical
and X-ray wavelengths (e.g. Urry \& Padovani 1995); these 
objects usually show weak or non-existent optical
emission lines. 

BL Lacs, together with Flat Spectrum Radio Quasars (FSRQs), are
called blazars. Blazars exhibit characteristics indicative
of relativistic beaming (Padovani \& Urry 1990) 
and seem to cover a wide range of polarization, strength
of the emission lines, and position of the synchrotron component
in the spectral energy distribution (SED) (Padovani \& Urry 2001).
An important distinction in terms of the SED was introduced
in BL Lacs by Padovani \& Giommi (1995); these
authors separated the objects emitting most of their synchrotron power
at high (UV--to-X-ray) and low (far-IR--to--near-IR) frequencies,
introducing the terminology of ``high-energy peaked BL Lac'' (HBL)
and ``low-energy peaked BL Lac'' (LBL) currently in use. Generally,
X-ray selected BL Lacs tend to peak at high energies and display
less extreme properties than the radio-selected ones, which peak
at low energies. It was suggested (Fossati et al. 1998;
Ghisellini et al. 1998) that the spectral properties and the
luminosity are anti-correlated, such that  high luminosity objects are
LBLs and low luminosity objects are HBLs.

In recent years there has been a growing interest in the
lower luminosity sources (e.g. Falcke 2001; Rector et al.
1999), as well as in those objects displaying
intermediate properties between HBLs and LBLs (Bondi et al. 
2001). Such objects are included in the 200 mJy sample 
of FRS sources defined by March\~a et al. (1996). 

March\~a's et al. sample of low luminosity, core dominated 
radio sources was selected with the purpose of finding low 
luminosity BL~Lacs and investigating the differences between 
such objects and other FRS sources.
Included in this sample were a number of galaxies which, 
although radio selected together with BL~Lacs and objects 
displaying a Seyfert-like emission spectrum, and following 
exactly the same criteria, are optically very different 
from them. In fact, they do not show any conspicuous signs 
of activity, the main contribution to their optical 
spectra being of stellar origin; such objects overlap with the 
ones dubbed PEGs (for Passive Elliptical Galaxies) described 
in Ant\'on (2000).  These could be objects 
related to BL~Lac phenomena but observed at larger angles to 
the jet;  it is also possible 
that some of the galaxies are ``hidden'' BL Lacs, whose 
nuclear emission is swamped by the host 
galaxy starlight (Dennett-Thorpe \& March\~a 2000). 

Among the galaxies displaying such characteristics, we have 
selected 19 sources for our studies; these objects are listed 
in Table~1 of a companion paper (Serote Roos \& 
Gon\c{c}alves 2003, hereafter Paper~I). Our purpose  was:
%Our study focus on a set of 19 such galaxies (listed in Table~1 
%of Paper~I) and has two  main goals:

{\it (i)}  to study the stellar content of LL FRS nuclei and to
test stellar population synthesis as a method allowing to reveal
any hidden optical emission-line features (these go usually
undetected, or are very weak, due to the strong dilution induced
by the stellar continuum);

{\it (ii)} to investigate the nature of the nuclear
emission regions and excitation mechanisms at the origin of the
weak emission lines present in the spectra of our objects.

\vspace{0.5mm} 
In Paper~I, we addressed the first  point; we
discussed the nuclear stellar populations and host properties of
the objects in our sample.  
In this paper we focus our attention on the nuclear emission-line 
spectra of LL FRS sources; these spectra are presented in Sect.~2. 
In Sect.~3 we introduce the line profile fitting method and describe 
how it has been applied to the data. In Sect.~4 we give some notes 
on the studied objects.
In Sect.~5 we discuss the implications of our findings, and in
Sect.~6 we summarize the results and conclusions of this work.

\section{The emission-line spectra}

\begin{figure*} 
\begin{center} 
\resizebox{!}{23cm}{\includegraphics{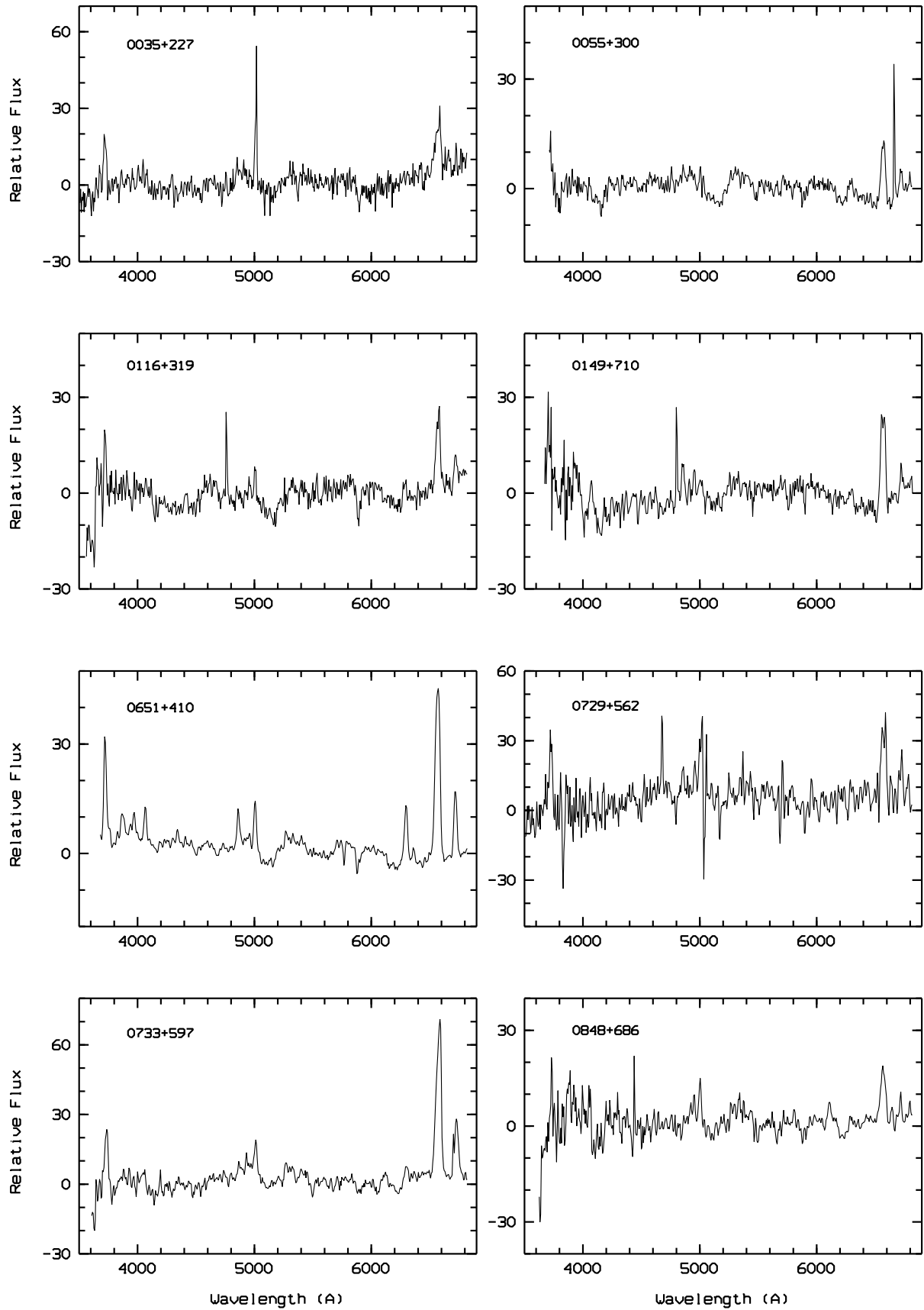}} 
\end{center} 
\caption{\label{em_line_spectra} 
Rest frame emission-line spectra obtained after correction 
of the stellar contribution.   
} 
\end{figure*}

Even though weak emission lines can be detected in the galaxies
belonging to our sample (see Fig.~1 in Paper~I), their optical 
spectra are clearly dominated by the stellar continu\-um  
which heavily dilutes the features associated 
with the active nucleus, thus making it difficult to access 
their true spectroscopic properties (profile, line width, 
intensity, etc.). 
Aware of this problem, we have calculated the stellar contribution 
to each spectrum and removed it from the data (see Paper~I),  
therefore obtaining a new set of 19 emission-line spectra; these are  
given in Fig.~\ref{em_line_spectra}.

We observe emission lines in all of the objects in 
Fig.~\ref{em_line_spectra}. 
Although emission lines are generally associated with   
the presence of an AGN, line emission is also frequently  
found in ``normal'' ellipticals (Goudfrooij 1999) and can be  
related to phenomena other than an active nucleus, e.g.:   
photoionization by very hot Wolf-Rayet stars (Warmers), hot O stars,  
or old post-AGB stars (Terlevich \& Melnick 1985; Filippenko \&  
Terlevich 1992; Binette et al. 1994), cooling accretion flows  
(Heckman 1981; Voit \& Donahue 1997), shock heating through cloud  
collisions induced by accretion, galaxy interactions, mergers or  
starburst-driven winds (Fosbury et al. 1978; Fosbury \& Wall 1979;   
Dopita \& Sutherland 1995, 1996; Alonso-Herrero et al. 2000), etc.   
 
In this study, we favour the AGN hypothesis and interpret the  
observed weak emission lines as the optical signature of an  
active nucleus. In support are the AGN-like radio characteristics  
of our sample,  namely a core-dominated emission and a flat radio  
spectrum, as well as the detection of an X-ray compact core in a 
few objects. 
It is intuitive to assume that the same central engine could be  
at the origin of the observed weak optical emission lines.  
These lines carry valuable
information on the line-emitting regions, ionizing continuum and
therefore, on the physical pro\-perties of the nuclei. In order to
extract these pieces of information from the spectra, a careful
modelling of the emission-line features and continuum is required.

\section{Line profile fitting and analysis}
All our emission-line spectra were fitted in terms of Gaussian
profiles thanks to a special routine, based on the least squares 
method, developed by E.\,J. Zuiderwijk, M.-P. V\'eron-Cetty and 
P. V\'eron. A full description of the method is given in 
Gon\c{c}alves (1999). 
In the present study, the emission lines \ha, \nii$\lambda$\,6548, 6584,
\sii$\lambda$6716, 6731 and \oi$\lambda$\,6300, 6363 (or \hb\ and
\oiii$\lambda$\,4959, 5007) were fitted by one  set of seven (or
three) Gaussian components; the width and redshift of each
component in a set were taken to be the same, as a result of the
lines being formed in the same region; this means that, in
addition to the line intensities, the free parameters for each set
of lines are one width and one redshift. The intensity ratios of
the \nii$\lambda$\,6548, 6584, \oiii$\lambda$\,4959, 5007 and
\oi$\lambda$\,6300, 6363 lines were set to their theoretical
values (3.00, 2.96 and 3.11, respectively; Osterbrock 1974). 

%------------------------------------ Fig. 1  ----------------------------- 
\addtocounter{figure}{-1} 
%------------------------------ Fig. 1 (cont) ---------------------------- 
\begin{figure*} 
\begin{center} 
\resizebox{!}{23cm}{\includegraphics{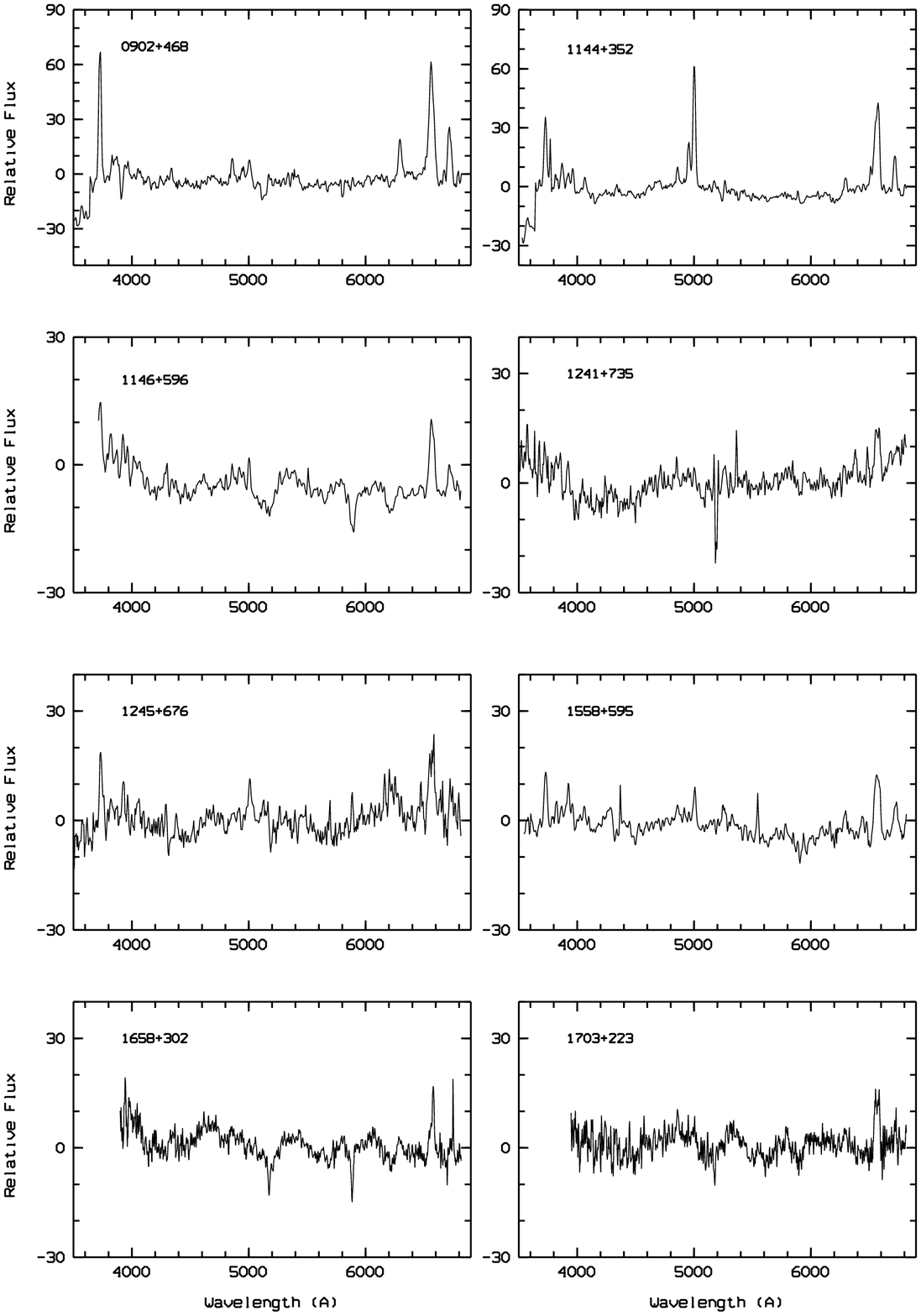}} 
\end{center} 
\caption{Rest frame emission-line spectra (cont.).} 
\end{figure*} 
\addtocounter{figure}{-1} 
%-------------------------------- Fig. 1 (end) --------------------------- 
\begin{figure} 
\begin{center} 
\resizebox{8.5cm}{!}{\includegraphics{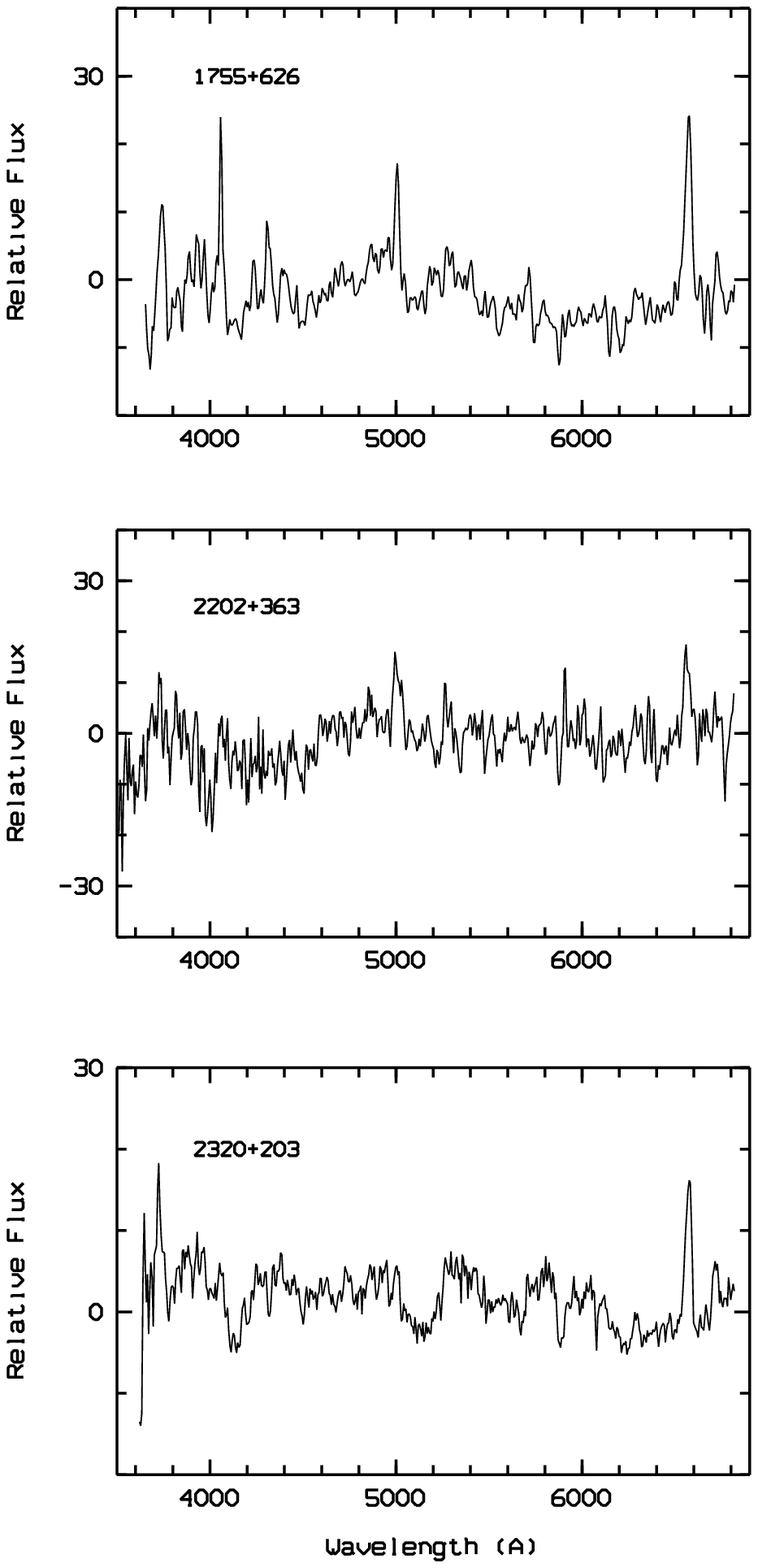}} 
% \resizebox{17cm}{!}{\includegraphics{fig6_nova_crop.ps}} 
\end{center} 
\caption{Rest frame emission-line spectra (end).} 
\end{figure} 
%----------------------------------------------------------------------- 
 
The emission lines used in our models were 
selected in function of their applicability in 
diagnostic diagrams. By plotting specific line ratios 
in a 2-dimensional diagram, we are able to identify the type 
of excitation mechanism at work in the galaxy nucleus; in 
particular, the use of the Baldwin et al.
(1981) or Veilleux \& Osterbrock (1987) diagnostic diagrams allows
to classify the nuclear emission-line regions of most galaxies
into one of three categories: nuclear \hii\ regions or starbursts,
Seyfert 2 galaxies and LINERs. Such diagrams make use of
conspicuous optical emission lines, such as \ha, \hb,
\nii$\lambda$6584, \oiii$\lambda$5007, \sii$\lambda$6717, 6713,
\oii$\lambda$3727 and \oi$\lambda$6300. 

\begin{figure*}[t]
%\begin{centre}
%% \psfig{figure=fig1.ps,height=23cm,width=18cm}
\resizebox{18cm}{!}{\includegraphics{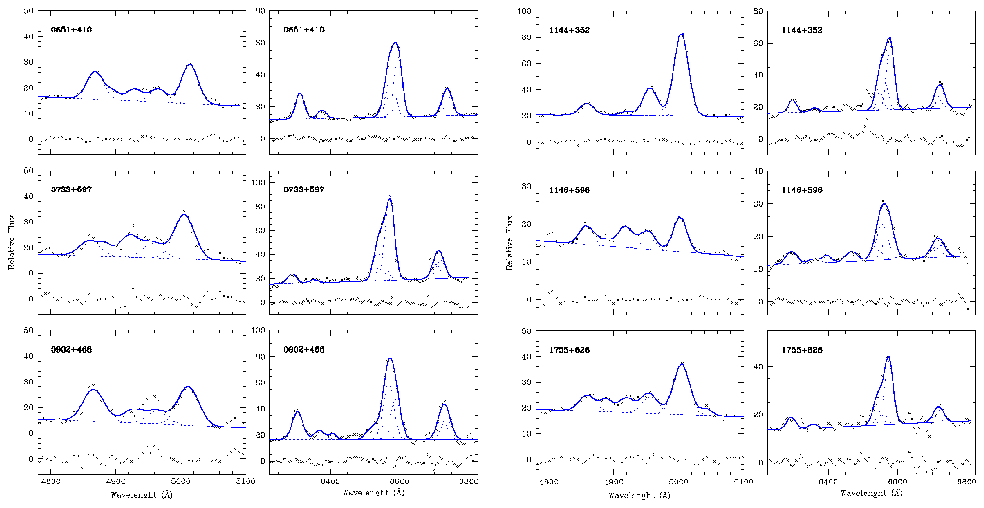}}
%\end{centre}
\caption{\label{fits} Best fitting results for six
nuclear emission-line regions; for each object, the blue
spectral region is shown on the left panel and the red
spectral region on the right panel. The data points are
represented by crosses; the  solid line is the best fit; 
the dotted lines show the individual components; the 
lower crosses  give the residuals. Both the 
data and the models were shifted upwards by an arbitrary amount 
for the sake of clarity. The residuals give an appreciation of 
the fit quality; in this figure, they are of the same 
order of magnitude as the noise in the spectra.} %\vspace*{2mm}
\end{figure*}

We have modeled the emission lines as explained above. In a few cases, 
additional Gaussians were added to fit iron blends, or other weak 
features present in the spectra; this did not affect the measured 
intensities of the lines of interest (i.e. the lines used to build
the diagnostic diagrams) but contributed to reduce the overall
scatter in the fit. Figure~\ref{fits} shows 
the best fitting models for six objects in our sample; the parameters 
resulting from the line profile fitting analysis are given in 
Table~\ref{fit_results}.  

\begin{figure*}[t]
\resizebox{12cm}{!}{\includegraphics{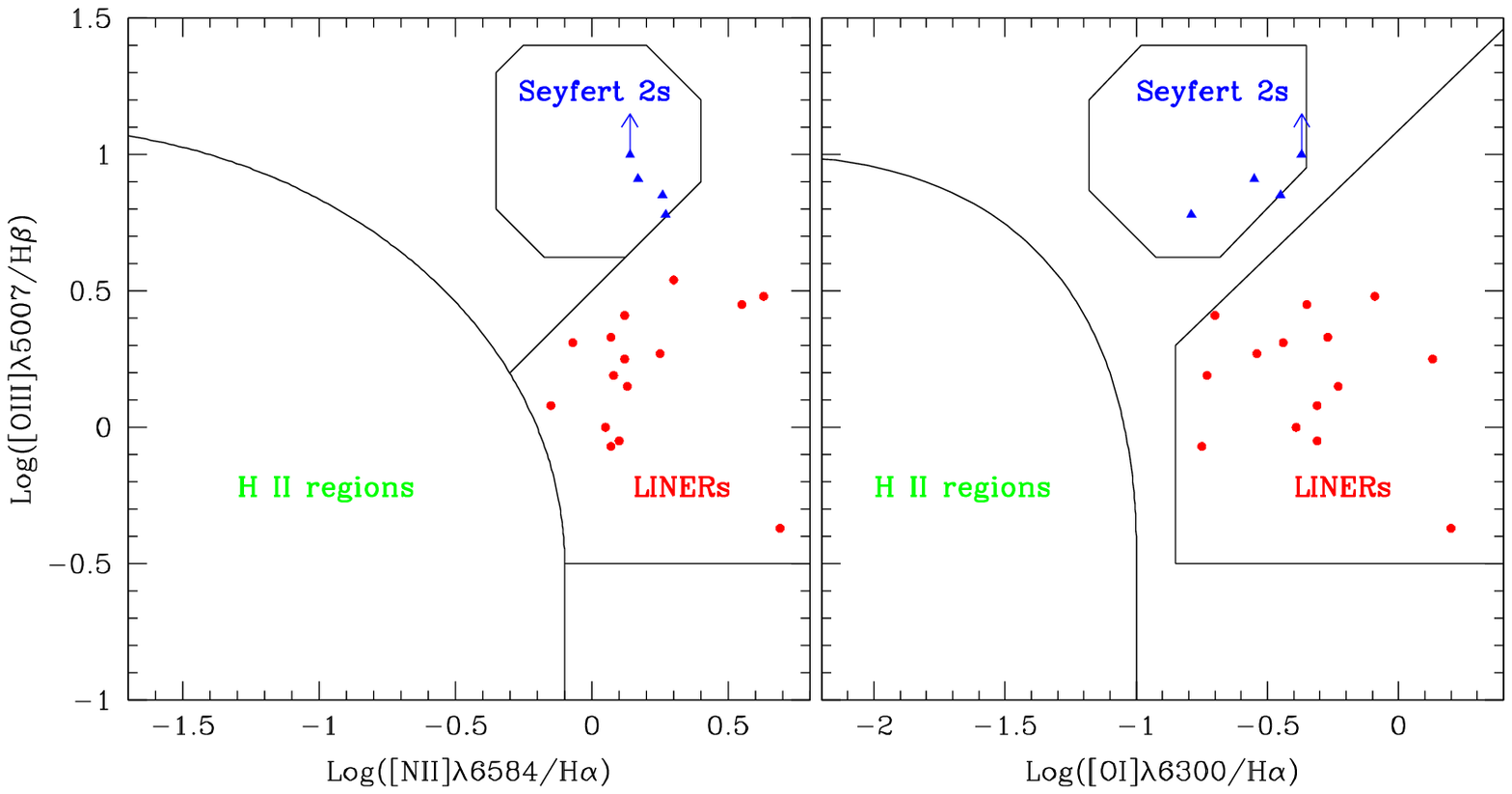}} \hfill
\parbox[b]{55mm}{
\caption{ Diagnostic diagrams showing the logarithm of \oiii\roiii\
{\it{vs.}} \nii\rnii\ and \oiii\roiii\ {\it{vs.}} \oi\roi. Most
of the objects in our sample are located in the LINER region
(represented as dots). Only a few nuclei display a
higher excitation spectrum, occupying the same location as Seyfert
galaxies (represented as triangles). The estimated errors are of the 
order of 0.10 and 0.13 in the x-axis (log\nii$\lambda$6584/\ha\   
and log\oi$\lambda$65300/\ha, respectively) and  0.11 in the 
y-axis.\vspace{8mm} }
\label{dd} }
\vspace{2.5mm}
\end{figure*}

The line ratios  \oiii$\lambda$5007/\hb,
\nii$\lambda$6584/\ha\ and \oi$\lambda$65300/\ha, 
used in the Veilleux \& Osterbrock (1987) diagnostic 
diagrams, take full advantage of the physical distinctions 
between the various types of objects and minimize the effects 
of redde\-ning correction and calibration errors. We have  
calculated these line ratios; the obtained values, given 
in Table~\ref{fit_results}, were used to build the 
diagnostic diagrams shown in Fig.~\ref{dd}. 
The line-ratios seem to be consistent, providing 
the same result on both diagnostic diagrams. According to 
these diagrams, 79\% (15/19) of the objects
harbour a LINER; the remaining nuclei (0035$+$227,
0116$+$319, 0848$+$686 and 1144$+$352) display a higher-ionization,
Seyfert-like spectrum. 

In this study, the mean error in the measurement of the 
\nii\ and \oiii\ lines intensity is of the order of 13\%, in the 
\hb\ and \oi\ lines it goes up to $\sim$24\%\ and in the \ha\ 
line it is  $\sim$19\%. Usually, no error bars are represented in 
diagnostic diagrams; however, an estimation of the errors 
can be calculated based on the previously given values, 
being of the order of 0.10 and 0.13 in the x-axis 
(log\nii$\lambda$6584/\ha\  and log\oi$\lambda$65300/\ha, 
respectively) and  0.11 in the y-axis 
(log\oiii$\lambda$5007/\hb) in Fig.~\ref{dd}. The 
classification is therefore robust. 

\section{Notes on individual objects}

{\bf 0035$+$227} is a radio-source detected at 5 GHz
(Griffith et al. 1990); %1990ApJS...74..129G
it has been studied at higher frequencies  by Dennett-Thorpe \&
March\~a (2000) and more recently at 1.4 GHz (Dennett-Thorpe,
private communication). This source has a steep spectral index
between 8 and 43~GHz (\a843) and a relatively low radio
polarization; this suggests that the object is not highly boosted,
as we would expect from a BL~Lac. Following Dennett-Thorpe \&
March\~a (2000), this could be a Compact Symmetric Object (CSO).
Our spectroscopic analysis points towards a nuclear region with
characteristics reminiscent of those of Seyfert~2s; such an
optical spectrum is more common amongst double-lobed, FR~II
radio-galaxies.

\medskip
{\bf 0055$+$300} is an FR~I radio-galaxy with an asymmetric
two-sided jet and a prominent core; its optical counterpart is the cD
galaxy NGC~315, at $z = 0.017$. This source has been extensively
studied using the VLA and VLBI ({\it e.g.} Venturi et al. 1993;
Cotton et al. 1999; Dennett-Thorpe \& March\~a 2000) and has been
detected at X-ray (Worrall \&  Birkinshaw  2000; Terashima et al. 2002) 
and infrared wavelengths (Golombek et al. 1988). 
Spectroscopic observations by Ho et al.(1995) %1995ApJS...98..477H
suggest the presence of a LINER with a weak broad \ha\ component.
We confirm the low ionization nature of this object, but were not
able to detect any broad Balmer line; however, such a component
cannot be excluded by our data. The detection of a broad component
would be in agreement with the other AGN-like characteristics of
the object, namely the presence of a radio jet and a compact,
unresolved, X-ray core. Although a small-scale cooling flow may be
present in NGC~315 (Worrall \&  Birkinshaw  2000), we interpret
the observed low ionization spectrum as the optical signature of a
low luminosity AGN.

%------------------------------------ Fig. 2 -----------------------------
%------------------------------ Fig. 2 (end) ----------------------------

\medskip
{\bf 0116$+$319} or 4C 31.04 is a variable radio-source with a
double-lobed radio structure; it shows the presence of a complex
\hi\ absorption (Conway 1996), which hides one of the mini-lobes
and partially covers the other. The core source was identified
with a faint component between the two extended lobes; no jet-like
structures were observed either
side of the core (Giovannini et al. 2001).  %2001ApJ...552..508G
From the steep \a843 value, Dennett-Thorpe \& March\~a (2000)
suggest that 4C 31.04 is not highly boosted; also its
symmetric structure and core-to-total radio power ratio
suggest that this source is near the plane of the sky
(Giovannini et al. 2001). 4C 31.04 is classified as a
low-redshift CSO  (Giovannini et al. 2001; Cotton et al.
1995). %2001ApJ...552..508G
This object is associated with a galaxy pair (MCG 5$-$4$-$17
and 5$-$4$-$18) at $z = 0.060$. Spectroscopic observations
by  Gelderman \& Whittle (1994) %1994ApJS...91..491G
reveal the presence of \ha, \nii\,$\lambda$6548, 6584 and
\oi\,$\lambda$6300 in emission, although
the quality of their spectra does not allow to classify the object
easily. Our data seem to indicate the presence of a nuclear
emission region with properties similar to those of Seyfert~2
galaxies.

\medskip
{\bf 0149$+$710} is a core dominated radio galaxy at
$z = 0.023$. It has been observed at 1.4 GHz by Lara et
al. (2001); %2001A&A...370..409L
its radio structure shows a strong asymmetry, with a wide jet in
the W-NW direction and a fainter component in the opposite
direction, probably related to the counter-jet. At 5 GHz, the VLBA
image shows a bright core and a collimated jet (Bondi et al.
2001). Thanks to a relatively high optical polarization (3.3\%)
and a \caii\ break contrast smaller than 0.4, this object has been
classified as a BL~Lac candidate by March\~a et al. (1996); in
addition, it shows a high radio polarization and a flat 8.4--43
GHz radio spectrum. Dennett-Thorpe \& March\~a (2000) suggest this
could be a ``hidden'' BL~Lac whose non-thermal continuum is being
swamped by starlight. Our spectroscopic analysis, carried on the
nuclear data after correction for the stellar contribution,
reveals a low ionization spectrum typical of LINERs. Detected in
the X-rays by ROSAT (Voges et al. 1999), 
this source is probably a low luminosity AGN observed close to the 
jet. 

\medskip
{\bf 0651$+$410} is a radio source associated with the elliptical
galaxy Zw 204$-$27. It shows a convex radio
spectrum, peaking at frequencies above a few GHz; this source is
thus a High Frequency Peaker, or
HFP (Dallacasa et al. 2000). %AA 363,887-900(2000)
However, it has been included by Marecki et al. (1999)
% Marecki A., Falcke H., Niezgoda J., et al., 1999, A&AS 135, 273
in their GPS (GHz-Peaked Spectrum) candidate list. This source
has a  break contrast $\sim$0.47 and an optical polarization
below the 2\% level;  due to its  borderline properties,
March\~a et al. (1996) could not classify this object. 0651$+$410
could be a weak BL Lac diluted by the starlight of the host
galaxy, as well as a flat radio spectrum  galaxy with a naturally
low break contrast. Spectroscopic observations by
Merighi et al. (1991) % 1991A&AS...89..225M
reveal the presence of \ha, \nii\,$\lambda$6548, 6584 and
\oi\,$\lambda$6300 in emission. Analysis of our nuclear
spectrum, corrected for the host galaxy starlight, shows the
presence of a LINER.

\medskip
{\bf 0729$+$562}  is a steep \a843\ radio source displaying
evidence of variability at 1.4 GHz (Dennett-Thorpe \&  March\~a
2000); although a number of BL Lacs and BL Lac candidates show
variability at radio wavelengths, this source does not seem to be
a BL~Lac object and remains intriguing. VLBI observations at 5 GHz
(Bondi et al. 2001) show that the radio source is rather
collimated and presents a series of bright and relatively compact
knots; however, the image does not allow us to decide whether the
object is one- or two-sided, nor to identify the core; the best
candidate, based on observations at 1.6 GHz, seems to be the
southernmost component. 0729$+$562  has been identified with a
weak-lined galaxy at $z = 0.107$. Our spectroscopic data suggests
the presence of a Low Ionization Nuclear Emission-line Region.

\medskip
{\bf 0733$+$597} is a compact radio source (Patnaik 1992) 
identified with an S0 galaxy at $z = 0.040538$. Observations with
the VLBI (Taylor et al. 1994) revealed a core-jet structure with a
faint counter-jet to the south; this source has been included in
the extragalactic radio-jets catalogue of
Liu \& Zhang (2002). %A&A...381..757L %
Following Dennett-Thorpe  \& March\~a (2000), 0733$+$597
would constitute a good beamed object candidate.
Analysis of our optical data reveals the presence of a
LINER, which is most probably associated with a low
luminosity AGN. The detection of this source at X-ray
wavelengths  with ROSAT (Voges et al. 1999) comes
in support of this interpretation.

\medskip
{\bf 0848$+$686} is a radio source with relatively flat 8.4--43
GHz radio spectrum and high radio polarization ($\sim$2.5\%); its
optical polarization is lower than 1\%. Our spectroscopic
analysis, carried out on the nuclear data after correction for the
stellar contribution, reveals the presence of a Seyfert-like
nuclear emission region. At the present time, we do not dispose of
additional pieces of information supporting the presence of an AGN
in this object. X-ray observations could contribute to enlighten
the situation and provide confirmation of the AGN nature of this
source. With a contrast value of 0.46 (March\~a et al. 1996), this
could then be another example of a ``hidden'' BL~Lac whose
non-thermal continuum is being swamped by the host galaxy
starlight. For Caccianiga et al. (2002), however, this is just
another Passive Elliptical Galaxy (PEG).

\medskip
{\bf 0902$+$468} is a compact radio source identified with a
galaxy at $z = 0.0848$. Its optical polarization value is less
than 1.2\%, and the radio polarization is inferior to 0.99\%
(March\~a et al. 1996; Dennett-Thorpe \& March\~a 2000). For
Caccianiga et al. (2002), this could be a narrow emission-line
galaxy; our analysis of the nuclear spectrum, corrected for the
stellar contribution, suggests that this object harbours a Low
Ionization Nuclear Emission-line Region.

\medskip
{\bf 1144$+$352} is a low power radio-galaxy showing variable
radio polarization at 8 GHz (Dennett-Thorpe \& March\~a 2000);
also its arcsecond radio core is variable.
This object has been identified with the Zwicky galaxy Zw~186.48 
%at $z = 0.0630$ 
and is a known GPS source (Snellen et al. 1995).
1144$+$352 was extensively studied by several authors, using
different instruments; VLA, MERLIN and VLBI data (Giovannini et
al. 1999) show a complex structure over a broad range of
physical scales (1 pc -- 1 Mpc) and confirm previous suggestions
of superluminal motion in this source. At parsec resolution,
Giovannini et al. (1999) identified the core source and two-sided
jets, very asymmetric in shape and properties; from these
observations, they computed a jet orientation with respect to the
line-of-sight of 25\degr; this is in agreement with the source
properties, intermediate between an FR~I galaxy and a BL Lac
object. 
This source was considered a BL~Lac candidate by March\~a et al.
(1996); in addition, it was detected at X-ray wavelengths by ROSAT
(Brinkman et al. 1995). All these pieces of evidence suggest that
this could, indeed, be a ``hidden'' BL~Lac, whose non-thermal
continuum is swamped by starlight, or an object with a jet at larger
angles to the line-of-sight. Analysis of the optical spectrum of
1144$+$352 suggests the presence of a high ionization nuclear
emission region reminiscent of those observed in Seyfert nuclei.
This is in agreement with the other AGN-like characteristics of
the object.

%------------------------------- Table 6 - Fitting profile analysis results ---
% \newpage
% \onecolumn
%\begin{sidewaystable}
\begin{table*}
\centering
%\small{
\caption{\label{fit_results} 
Results of the line profile fitting analysis.  Col. 1 gives the name of the 
object, col. 2 the adopted redshift, cols. 3 and 6 the velocities for each 
set of components measured on the blue and red spectra, respectively, and 
de-redshifted using the redshift given in col.~2; cols. 4 and 7 the 
corresponding {\it FWHM} (uncorrected for the instrumental broadening), 
cols. 5, 8  and 9 the intensity ratios \oiii\roiii, \nii\rnii\ and 
\oi\roi, respectively; n.d. means we could not detect one of the lines 
(intensity $\sim 0$).  
%In col. 10 we give the velocity difference between 
%the blue and red systems, and 
In col. 10 we give the spectroscopic classification 
of the nuclear emission-line region resulting from the diagnostic 
diagrams.}
%\begin{flushleft}
%\begin{tabular}{ll|rlr|rlrr|rc}
\begin{tabular}{clrrrrrrrc}
\hline
\hline
 Name   &  \verb+  +$z$       &   $V\; \; \; \; \; \;$         & 
$FWHM\/$  &   
%\underline{$\lambda$5007$\:$}                      & 
\underline{\oiii}$\;$        &
$V\; \; \; \; \; \;$         & $FWHM\/$        & 
%\underline{$\lambda$6583$\:$} &
\underline{\nii}      &
%\underline{$\lambda$6300$\:$}               
\underline{\oi}$\;\;\;\;$       
% & $\Delta V \verb+  + $         
&    Class. \\
        &                     &  (\kms)                        &
 (\kms) & \hb$\;\;$                                            &
 (\kms) & (\kms)              & \ha$\;$     & \ha$\;\;\;\;$    &
%                                (\kms)        & 
               \\
\hline

  0035$+$227      &    0.097     &    $-$200 \verb+  + &  1078 \verb+ +     & 
% 0.78\verb+ +    
  6.09\verb+ +    &    $-$998 \verb+  +                &  976 \verb+ +      & 
% 0.27            &    $-$0.77 \verb+ +     
  1.87            &    0.17 \verb+ +                   &  
%798 \verb+  +     & 
  Seyf 2         \\

  0055$+$300      &    0.017     &    $-$111 \verb+  + &   908 \verb+ +     & 
% 0.19\verb+ + 
  1.53\verb+ +    &    24 \verb+  +                    &   1285 \verb+ +    & 
% 0.08            &    $-$0.73 \verb+ +
  1.20            &    0.19 \verb+ +                   &   
%$-$135 \verb+  + & 
  LINER          \\

  0116$+$319      &    0.060     &    $-$682 \verb+  + &   1506 \verb+ +    & 
% 0.91\verb+ +  
  8.22\verb+ +    &    $-$945 \verb+  +                &   887 \verb+ +     & 
% 0.17            &    $-$0.55 \verb+ + 
  1.47            &    0.28 \verb+ +                   &   
%263 \verb+  +    & 
  Seyf 2         \\

  0149$+$710      &    0.023     &    243 \verb+  +    &   1463 \verb+ +    & 
% $-$0.07\verb+ + 
  0.85\verb+ +    &    786 \verb+  +                   &   1042 \verb+ +    & 
% 0.07            &    $-$0.75 \verb+ + 
  1.17            &    1.18 \verb+ +                   &   
%$-$543 \verb+  + & 
  LINER          \\

  0651$+$410      &    0.022     &    408 \verb+  +    &  1846 \verb+ +     & 
% 0.15\verb+ +        
  1.40\verb+ +    &    552 \verb+  +                   &  1290 \verb+ +     & 
% 0.13            &    $-$0.23 \verb+ +                  
  1.36            &    0.59 \verb+ +                   &  
%$-$144 \verb+  +  & 
  LINER          \\

  0729$+$562      &    0.107     &    336 \verb+  +    &  1728 \verb+ +     & 
% 0.41\verb+ +                 
  2.60\verb+ +    &    663 \verb+  +                   &  782 \verb+ +      & 
% 0.12            &    $-$0.70 \verb+ +
  1.31            &    0.20 \verb+ +                   &  
%$-$327 \verb+  +  & 
  LINER          \\

  0733$+$597      &    0.041     &    $-$101 \verb+  + &  2273 \verb+ +     & 
% 0.45\verb+ +              
  2.84\verb+ +    &    $-$519 \verb+  +                &  1432 \verb+ +     & 
% 0.55            &    $-$0.35 \verb+ +                  
  3.53            &    0.45 \verb+ +                   &  
%418 \verb+  +     & 
  LINER          \\

  0848$+$686      &    0.041     &   $-$1004 \verb+  + &  1744 \verb+ +     & 
% $\ga 1$ \verb+ +  
$\ga10.\;\;\;\;\:\,$ & $-$1242 \verb+  +               &  1490 \verb+ +     & 
% 0.14            &    $-$0.37 \verb+ +                  
  1.38            &    0.42 \verb+ +                   &   
%238 \verb+  +    & 
  Seyf 2         \\

  0902$+$468      &    0.085     &    70 \verb+  +     &  2327 \verb+ +     & 
% 0.08\verb+ +    
  1.19\verb+ +    &    225 \verb+  +                   &  1584 \verb+ +     & 
% $-$0.15         &    $-$0.31 \verb+ +                
  0.71            &    0.49 \verb+ +                   &  
%45 \verb+  +      & 
  LINER          \\

  1144$+$352      &    0.064     &    $-$189 \verb+  + &  1588 \verb+ +     & 
% 0.85\verb+ +             
  7.06\verb+ +    &    $-$189 \verb+  +                &  1200 \verb+ +     & 
% 0.26            &    $-$0.45 \verb+ +                  
  1.81            &    0.36 \verb+ +                   &  
%0 \verb+  +       & 
  Seyf 2         \\

  1146$+$596      &    0.011     &    $-$293 \verb+  + &  1476 \verb+ +     & 
% 0.31\verb+ +                   
  2.03\verb+ +    &    $-$331 \verb+  +                &  1731 \verb+ +     & 
% $-$0.07         &    $-$0.44 \verb+ +                
  0.84            &    0.36 \verb+ +                   &  
%38 \verb+  +      & 
  LINER          \\

  1241$+$735      &    0.075     &    291 \verb+  +    &  1455 \verb+ +     & 
% $\sim$0 \verb+ + 
  1.01\verb+ +    &    1059 \verb+  +                  &  1192 \verb+ +     & 
% 0.05            &    $-$0.39 \verb+ +                
  1.13            &    0.41 \verb+ +                   &  
%$-$768 \verb+  +  & 
  LINER          \\

  1245$+$676      &    0.107     &     741 \verb+  +   &  2000 \verb+ +     & 
% 0.54\verb+ +    
  3.51\verb+ +    &    462 \verb+  +                   &  1036 \verb+ +     & 
% 0.30            &    $\sim$0 \verb+ +   
  1.99            &    n.d. $\;\;$                     &  
%279 \verb+  +     & 
  LINER?         \\

  1558$+$595      &    0.060     &    $-$498 \verb+  + &  1606 \verb+ +     & 
% 0.38\verb+ +    
  2.40\verb+ +    &    $-$751 \verb+  +                &  1496 \verb+ +     & 
% 0.0\verb+ +7    &    $-$0.27 \verb+ + 
  1.18            &    0.53 \verb+ +                   &  
%253 \verb+  +     & 
  LINER          \\

  1658$+$302      &    0.036     &     549 \verb+  +   &  1664 \verb+ +     & 
% $-$0.37\verb+ + 
  0.42\verb+ +    &    1203 \verb+  +                  &  958 \verb+ +      & 
% 0.69            &    0.20 \verb+ + 
  4.91            &    1.59 \verb+ +                   &  
%$-$654 \verb+  +  & 
  LINER          \\

  1703$+$223      &    0.050     &    $-$877 \verb+  + &  1875 \verb+ +     & 
% $-$0.05\verb+ +                                      
  0.88\verb+ +    &    $-$1418 \verb+  +               &   915 \verb+ +     & 
% 0.10            &    $-$0.31 \verb+ +
  1.27            &    0.48 \verb+ +                   &   
%514 \verb+  +    & 
  LINER          \\

  1755$+$626      &    0.028     &    $-$153 \verb+  + &  1795 \verb+ +     & 
% 0.48\verb+ +    
  3.04\verb+ +    &    $-$386 \verb+  +                &  1346 \verb+ +     & 
% 0.63            &    $-$0.09 \verb+ +                  
  4.29            &    0.81 \verb+ +                   &  
%233 \verb+  +     & 
  LINER          \\

  2202$+$363      &    0.075     &    795 \verb+  +    &   3027 \verb+ +    & 
% 0.25\verb+ +  
  1.79\verb+ +    &    921 \verb+  +                   &   1913 \verb+ +    & 
% 0.12            &    0.13 \verb+ +  
  1.33            &    1.35 \verb+ +                   &   
%$-$126 \verb+  + & 
  LINER          \\

  2320$+$203      &    0.039     &    228 \verb+  +    &   2017 \verb+ +    & 
% 0.27\verb+ + 
  1.86\verb+ +    &    678 \verb+  +                   &   1067 \verb+ +    & 
% 0.25            &    $-$0.54 \verb+ + 
  1.79            &    0.28 \verb+ +                   &   
%$-$450 \verb+  + & 
  LINER          \\

\hline
\end{tabular}
%\end{flushleft}}
%\end{center}
%\end{sidewaystable}
\end{table*}
%\newpage
%\twocolumn
%%%%%%%%%%%%%%%%%%%%%%%%%%%%%%%%%%%%%%%%%%%%%%%%%%%%%%%%%%%%%%%%%%%%%%%%%%%%%%%

\medskip
{\bf 1146$+$596} is a compact radio source for Condon \& Dressel
(1978). VLBI observations at 5 GHz (Wrobel et al. 1985) suggest
the presence of three components consistent with  an asymmetric
core-jet structure; further VLBI imaging of this source provided
morphological and spectral evidence leading to different
conclusions -- namely the presence of twin, parsec-scale jets. The
twin-jet kinematics requires that the jets are mildly relativistic
and oriented at $\sim 50\degr$ from the line-of-sight
(Taylor et al. 1998). %THE ASTROPHYSICAL JOURNAL, 498:619624, 1998
1146$+$596 has been included in the extragalactic radio-jets
catalogue of Liu \& Zhang (2002);  %A&A...381..757L %
the optical counterpart of this source is NGC~3894, a bright
galaxy at  $z =$ 0.01068 classified as an elliptical or 
an S0 (Nilson 1973). Although its
optical continuum is dominated by starlight, this galaxy's radio
and far-infrared continuum suggest the presence of an active
nucleus (Condon \& Broderick 1988). Spectroscopic observations
carried by Kim (1989) show the presence of a dust lane and ionized
gas along the galaxy's major axis; the gas kinematics are rather
peculiar, exhibiting non-circular motions. Analysis of our data
shows that the nuclear emission-line region has LINER
characteristics; such a low ionization spectrum could be
associated with a  low power AGN.

\medskip
{\bf 1241$+$735} is a core-jet radio source (Augusto et al. 1998)
identified with a bright galaxy at $z =$ 0.075. VLBI imaging by
Bondi et al. (2001) confirms the compact core and one-sided jet
and reveals an unusual radio morphology at  larger distances from
the core, where the jet starts bending and ends in a compact
component. There is no hint of a counter-jet or of any extended
emission on the side opposite to the main radio jet. This object
shows very high and variable radio polarization and a flat \a843\
index, which led Dennett-Thorpe \& March\~a (2000) to suggest this
could be an active nucleus with a jet  close to the
line-of-sight. Analysis of our spectroscopic data shows the
presence of a Low Ionization Nuclear Emission-line Region, which
is compatible with the presence of a low power AGN.
%Augusto et al. 1998 1998MNRAS.299.1159A

\medskip
{\bf 1245$+$676} is a giant radio-galaxy with FR II type
morphology (de Vries et al. 1997); it is also a  known GPS source.
At 1.4 GHz, its radio emission is core-dominated (Lara et al.
2001). This source has been identified with the galaxy VII~Zw 485
at $z= 0.107$. It displays a high contrast value  and
little radio polarization (Dennett-Thorpe \& March\~a 2000); in
consequence, this does not seem to be a good beamed object
candidate. Spectroscopic observations published by de Vries et al.
(2000) show the presence of \ha, \hb, \nii$\lambda$\,6584 and
\sii$\lambda$\,6717, 6731 in emission; our data, corrected for the
stellar contribution and modelled with a powerful line-deblending
method, confirm these detections as well as the detection of
\nii$\lambda$\,6548 in emission. However, the poor signal-to-noise
ratio of our spectrum does not allow a firm detection of the
\oi$\lambda$\,6300, 6363 lines; unfortunately, the spectrum
published by  de Vries et al. was not corrected for the
atmospheric absorption, and displays a deep H$_{2}$O absorption at
the wavelengths of interest,  making it impossible to check for
the presence of the \oi\ emission lines. Our LINER classification,
based on one diagnostic diagram alone, is therefore to be checked
through the analysis of better quality data. The presence of an
AGN in this source needs still to be confirmed.

%----------------------------------- fig 2 (DD) --------------------------
%--------------------------------------- end fig DD ----------------------

\medskip
{\bf 1558$+$595} is a radio source identified with an elliptical
galaxy at $z =  0.0602$. It displays a high contrast
value and little radio polarization (Dennett-Thorpe \& March\~a
2000); from its steep \a843\ index, these authors conclude that
this is not a  highly boosted object and suggest it could be a
CSO. VLBA images at 5 GHz show a compact triple-component
morphology (Bondi et al. 2001); it is not clear which component is
the core. NVSS maps reveal the presence of two slightly resolved
radio features, connected by fainter emission, plus another two
features to the East, with a two-sided jet visible (Denett-Thorpe,
private communication). Spectroscopic observations of this object
point towards the presence of a Low Ionization Nuclear
Emission-line Region. This is not supported by Caccianiga et al.
(2002), who classify 1558$+$595 as a simple PEG.

\medskip
{\bf 1658$+$302} or 4C~30.31, is a radio galaxy with  FR~I
type morphology; it shows a 32 kpc one-sided
jet embedded in diffuse emission (Gonzalez-Serrano et al. 1993). 
Capetti et al. (2000) revealed the presence, in this galaxy, of
a faint one-sided optical jet, co-spatial with the radio jet. 
Images obtained at 1.4 GHz show a small double with a compact
point source at some distance (Dennett-Thorpe, private
communication). This object's radio polarization is below the 2\%
level; the optical polarization value is even lower
(Dennett-Thorpe \& March\~a 2000; March\~a et al. 1996).
1658$+$302 does not seem to constitute a good blazar candidate.
Analysis of our spectrum suggests the presence of a LINER.

\medskip
{\bf 1703$+$223} is a radio source associated with an elliptical
galaxy at $z = 0.04977$. Although its optical polarization level
is inferior to 2\%, March\~a et al. (1996) have classified this
object as a BL Lac candidate based on the contrast value; this has
been recalculated by Dennett-Thorpe \& March\~a  (2000), whom no
longer classify 1703$+$223 as a beamed object. Observed at 1.4
GHz, this source displays two small compact doubles
(Dennett-Thorpe, private communication). Analysis of our
spectroscopic data, after correction for the stellar contribution,
shows the presence of a Low Ionization Nuclear Emission-line
Region.

\medskip
{\bf 1755$+$626} is a radio source identified with NGC~6521, a
galaxy in a cluster. It has a radio polarization value inferior to
2\% and a flat high frequency radio spectrum; based on these
characteristics, Dennett-Thorpe \& March\~a  (2000) have suggested
this could be a ``hidden'' BL Lac whose thermal continuum is
swamped by starlight. Analysis of our optical spectrum, after
correction for the stellar contribution, reveals the presence of a
LINER; this could be associated with a low power AGN. The X-ray
detection of this source by Zimmermann et al. (2001) supports this
hypothesis. 1755$+$626 is thus likely to host a beamed active nucleus. 

\medskip
{\bf 2202$+$363} is a compact radio source identified with
a galaxy at $z = 0.075$. Unresolved at 1.6 GHz, this
source has been observed with  the VLBA  at 5 GHz (Bondi et al.
2001) showing a single component, barely resolved, and no
indications of extended emission. This object shows variability
at 1.4 GHz (Dennett-Thorpe, private communication); its radio
polarization level is inferior to 1.26\%, the optical polarization
being significatively lower (March\~a et al. 1996;  Dennett-Thorpe
\& March\~a  2000). Our spectrum shows the presence
of a Low Ionization Nuclear Emission-line Region, which could be
related to the presence of a low power AGN.

\medskip
{\bf 2320$+$203} is a compact radio source identified
with a galaxy at $z = 0.039$. This is another example
of an object displaying high radio polarization and a flat
\a843 index (Dennett-Thorpe \& March\~a 2000). Our spectroscopic
analysis reveals the presence of a Low 	onization Nuclear
Emission-line Region; this is most probably associated
with a low power AGN. Detection of this object at X-rays wavelengths
by Brinkmann et al. (1995) %1995A&AS..109..147B
comes in support of this hypothesis. 2320$+$203 is thus
likely to host an active nucleus observed close to the jet.

\section{Discussion}

\subsection{FRS Sources and Low Luminosity AGN}

Although usually displaying high luminosities, FRS
sources seem to span a large range in luminosity
and even jet power (Falcke et al. 2001). In the past 
few years there has been a growing interest in the 
low luminosity end of the distribution, 
with some studies focusing on low power sources
(Paper~I; Falcke 2001;  Falcke et al. 2001;
March\~a et al. 1996;  Rector et al. 1999).
Some of the above mentioned studies were driven by 
searches for the true nature of blazars and helped
to explore the limitations of the present
classification.

In low power sources, the emission coming from the host galaxy
(hot gas and starlight) is more prone to dilute the non-stellar
spectrum, therefore adding to the difficulty in studying the
optical properties of the objects. Fully aware of this fact, we
have revisited March\~a et al. data in search of additional
information on the nuclear region of LL FRS sources. 
Our results show that the majority of the studied nuclei
have LINER-type spectra. It is interesting to note that
these low luminosity sources display the same optical
properties as other low luminosity AGN (LL AGN), namely
a low ionization spectrum.

Is the LINER emission a signature of a low power AGN, or
is it being produced by other phenomena  (shocks, photoionization
by very hot stars, etc.)?  
The AGN hypothesis is  favored in our objects; in support, are the
AGN-like radio characteristics of the sources  as well as the
detection of an X-ray compact core in a few objects. Some of the
nuclei show a relatively high degree of radio and optical
polarization (see Table~1 of Paper~I), which supports their beamed
nature. Although most of the sources do not seem to be highly
boosted, they can nevertheless be powered by a low luminosity AGN
with a jet at larger angles to the line-of-sight. 

Our results show that there is a connection between LL~FRS sources
and low power AGN; other studies give support to our findings. For
instance, Nagar et al. (2000, 2002) and Falcke et al. (2000, 2001) 
have reported VLA and VLBA observations of a sample of
LL~AGN selected from the spectroscopic survey of Ho et al. (1995); 
spectroscopically, most of the
objects in this survey are classified as LINERs, with some low
luminosity and composite LINER-\hii\ objects also present (Ho et
al. 1997). Their observations have revealed that a large fraction
of the studied LINERs have flat spectrum compact radio cores,
similar to those found in many AGN; in the brightest cores it was
even possible to resolve the radio emission into jet-like
structures. These authors sustain that at least 50\%\ of low
luminosity AGN in the sample are accretion dominated, with the
radio emission presumably coming from the jets and/or an
advection-dominated accretion flow. In addition, the brightness
temperatures of the radio cores are consistent with the presence
of an active nucleus, giving additional support to the AGN-like
nature of these LINERs. 

Observing the cores of a sample of LL AGN,
such as LINERs, these authors found a number of BL Lac-like
sources. Our study followed the inverse path: studying a sample of
low luminosity FRS sources, known to contain a high fraction of
objects with blazar-like properties, we conclude that the large
majority display a low ionization spectrum, typical of LINERs.

\subsection{LINERs and ADAFs}
In LL AGN such as LINERs, it has  been suggested that radio cores
could be due to an advection-dominated accretion flow, or ADAF
(Narayan \& Yi 1994, 1995a,b; Fabian \& Rees 1995); an ADAF occurs
for small values of the accretion rate (less than 1\%\ of the
Eddington rate). 

There are estimations of the BH masses in BL Lacs and radio-loud 
quasars (e.g. Treves et al. 2002; Falomo et al. 2002; Kotilainen  
et al. 2002). BL Lac BH masses span a range 5\,10$^{7}$ -- 
10$^{9}$~$M_{\sun}$ (Falomo et al. 2002; Kotilainen et al. 2002 ); 
additional results
show that the velocity dispersions and BH masses of BL Lac objects
are similar to those obtained for low-$z$ radio-galaxies, in
agreement with the unified schemes for BL~Lacs and radio-galaxies;
also, the average BH masses in radio-loud quasars are a factor
$\sim$2 greater than those of BL Lacs. The BH masses can be used
to derive the Eddington luminosity  and this value can be used to
calculate the Eddington ratio, provided the total luminosity is
known. Treves  et al. (2002) estimated the Eddington ratio for the
two classes of objects; their values differ by two orders of
magnitude in the two classes.  This gives support to the
hypothesis that the accretion rate changes from largely
sub-Eddington in low luminosity, weak-lined sources, to
near-Eddington in high luminosity, strong-lined sources.

It has been suggested that the main difference between FR~I and
FR~II radio-galaxies, as well as between LINERs and Seyferts, 
lies in  the accretion rate (for a complete discussion, see 
the review by V\'eron-Cetty \& V\'eron 2000); in the former
objects, the accretion rate is small (and so is the $L/L_{\rm
Edd}$) and thus, an ADAF forms (Baum et al. 1995). There are
evidences for the presence of an ADAF in the radio source Sgr
A$^{*}$, which has a LINER spectrum (Yuan et al. 2002) and 
also in the FR~I radio-galaxy M\,87
(Reynolds et al. 1996). Moreover, LINERs show characteristics
which can be accounted for by the presence of an ADAF, like little
short-term X-ray variability, or the presence of double-peaked
profiles. 
Better quality data would allow to search for the possible presence 
of such double-peaked emission lines in the galaxies of our sample.

An advection-dominated accretion flow could be present 
in most BL Lacs and BL Lac-type objects. Cao (2002) studied a
sample of BL Lac objects and found a bi-modal distribution in
their \hb\ luminosity; this author suggests that standard thin
disks are associated with sources displaying $L_{\rm H\beta} >
10^{41}$ \ergs\ and BH masses in the range 10$^{8-10} M_{\sun}$.
In lower luminosity sources, like the ones in our sample, he
expects the standard thin disk to evolve into an advection-dominated 
accretion flow, forming a hot and thick disk. The lower
limit to the BH mass of these objects is in the range 1.66--24.5
10$^{8} M_{\sun}$, in agreement with what is observed in LINERs.

\subsection{The unifying scheme for BL Lacs and FR~Is}
Blazars seem to be associated with Fanaroff-Riley type I and II
radio-galaxies. According to the unifying scheme, FSRQs and FR~II 
radio-galaxies would be similar objects observed from different 
angles between the radio jet and the line-of-sight (Barthel 1989), 
with the former being the beamed version of the latter. 
In what concerns BL Lacs,
there is still some debate on the nature of the parent population;
some authors believe it to be the low luminosity, core dominated,
FR~I galaxies (Urry \& Padovani 1995; Urry et al. 1999, 2000;
Kollgaard et al. 1992), while others suggest FR~IIs may also
contribute (Murphy et al. 1993; Cassaro et al. 1999). A transition
population between beamed BL Lacs and unbeamed radio-galaxies has
not been detected (Rector et al. 1999); yet, a transition
population of low luminosity BL Lacs was predicted to exist in
abundance in X-ray selected samples, like the one studied by
Browne \& March\~a (1993).

A hot, radiatively inefficient accretion flow such as an ADAF
could contribute to suppress the thermal emission in blazar-like
objects, therefore allowing for the non-thermal radiation to
dominate over larger angles between the jet and the line-of-sight. As a
consequence, it would be possible to  observe some of the blazars'
properties without relativistic beaming taking place. 
Such objects could be linked to a transition population 
between beamed BL~Lacs and unbeamed FR~I galaxies. 
This could be the case for some of 
the objects in our sample, e.g. 1146$+$596 or 1144$+$352.

FR~I galaxies usually display an optical LINER spectrum
(Laing et al. 1994; Gon\c{c}alves 1999) unlike FR~IIs, which
show nuclear spectra of higher ionization (either Seyfert~1- or
Seyfert 2-like). PKS 2014$-$55 (Jones \& McAdam 1992; Simpson et
al. 1996) and S5~2116$+$81 (Lara et al. 1999) seem to be the
exception, displaying simultaneously a FR~I radio morphology and a
Seyfert-type spectrum. It would be interesting to follow these
objects with a long-term spectroscopic program and check for
variability in their spectra; detection of a LINER-type spectrum
could be an indication that the object has exhausted the gas near 
the black hole and will undergo an ADAF phase. 

\section{Summary and Conclusions}
In an effort to better understand low luminosity FRS sources, we
undertook the spectroscopic study of 19 objects selected from
March\~a's et al. (1996) 200~mJy sample. This sample was expected 
to contain a high fraction of objects observed at small angles to 
the jet,  like low luminosity BL Lacs and candidate BL Lacs, 
and includes objects displaying intermediate properties between
HBLs and LBLs.

Our study made use of stellar population synthesis as a tool to
recover the true spectroscopic characteristics of the nuclei; we
then performed a careful analysis of the resulting emission-line
spectra. The data were modelled in terms of Gaussian profiles and
the measured line ratios were used to identify the excitation
mechanisms at work in the nuclear emission-line regions by means
of diagnostic diagrams. The results of this analysis show that
79\%\ (15/19) of the objects harbour a Low Ionization Nuclear
Emission-line Region, or LINER.

The observed low ionization spectra are interpreted as the optical
signature of an active nucleus. 
Although most of the sources in our sample do not seem to be
highly boosted, they can nevertheless be powered by a low power
AGN with a jet  at larger angles to the line-of-sight. 
The stellar populations derived in Paper~I for these nuclei are in
agreement with those usually found in LINERs; also the black hole
masses reported for some LL FRS sources are compatible with the
presence of a LL AGN, such as a LINER. Our results support a 
unifying picture involving LL FRS sources and FR~I radio-galaxies,
which usually display a low ionization optical spectrum. The fact
that we observe a LINER-type spectrum in 79\%\ of the nuclei
suggests that the majority of the objects presented here could be
undergoing an ADAF phase. The remaining sources may still dispose
of enough gas around the black hole to fuel an optically thick,
geometrically thin, cool accretion disk; as a consequence, we
observe a higher-ionization, Seyfert-like spectrum.

Further observations, carried out at higher spectral resolution and
better signal-to-noise ratio, would allow us to better constrain the
spectroscopic properties of these nuclei and also to check for
variations in their line profiles; it would be particularly
interesting to pursue such a study on the few objects displaying
radio variability. Finally, better quality 
data would allow us to search for large Balmer line components and
double-peaked emission lines, typical of relativistic accretion
disks.

\bigskip
{\it Acknowledgements:} 
We thank S. Ant\'on for fruitful discussions and P. V\'eron for 
valuable comments.
A.\,C. Gon\c{c}alves and M. Serote Roos acknowledge support from
the {\it {Funda\c{c}\~ao para a Ci\^encia e a Tecnologia}}, Portugal,
under grants no. BPD/9422/02 and BPD/5684/2001.

% -------------------------------------------------------------------------

% ----------------------------------------------------------------------
\end{document}